\documentclass[journal]{IEEEtran}
\ifCLASSINFOpdf
\else
\fi
\usepackage{nomencl}
\usepackage{amssymb}
\usepackage{float}
\usepackage{graphicx}
\usepackage{subfigure}
\usepackage{etoolbox}
\usepackage{cite}
\usepackage{amsmath,amssymb,amsfonts}
\usepackage{amsmath,textcomp}
\usepackage{algorithmic}
\usepackage{textcomp}
\usepackage{xcolor}
\usepackage{multirow}
\usepackage{makecell}
\usepackage{booktabs}
\usepackage{tipa}
\usepackage{dsfont}
\usepackage{textcomp}
\usepackage{threeparttable}
\usepackage{threeparttable}
\usepackage{upgreek}
\usepackage{mathrsfs}
\usepackage{CJKutf8}
\usepackage{xcolor}
\makenomenclature
\begin{document}
\title{Joint Planning of Active Distribution Network and EV Charging Stations Considering Vehicle-to-Grid Functionality and Reactive Power Support}
\author{Yongheng~Wang,~\IEEEmembership{Student Member,~IEEE,}
        Xinwei~Shen,~\IEEEmembership{Senior Member,~IEEE,}
        Yan~Xu,~\IEEEmembership{Senior Member,~IEEE}
\thanks{Y. Wang and X. Shen are with Institute for Tsinghua Shenzhen International Graduate School, Tsinghua University. Y. Xu is with the School of Electrical and Electronic Engineering, Nanyang Technological University.}
\thanks{This work is supported in part by National Natural Science Foundation of China (No.52007123). Corresponding author: Xinwei Shen, email: sxw.tbsi@sz.tsinghua.edu.cn}
}
\maketitle
\begin{abstract}
This paper proposes a collaborative planning model for the active distribution network (ADN) and electric vehicle (EV) charging stations that fully considers the vehicle-to-grid (V2G) function and reactive power support of EVs in different regions. This paper employs a sequential decomposition method based on the physical characteristics of the problem, breaking down the holistic problem into two sub-problems for solution. Subproblem I optimizes the charging and discharging behavior of the autopilot electric vehicles (AEVs) using a mixed-integer linear programming (MILP) model. Subproblem II uses a mixed-integer second-order cone programming (MISOCP) model to plan the ADN and retrofit or construct V2G charging stations (V2GCS), as well as multiple distributed generation resources (DGRs). The paper also analyzes the impact of the bi-directional active-reactive power interaction of V2GCS on ADN planning. The presented model was tested in the 47 nodes ADN in Longgang District, Shenzhen, China, and the IEEE 33 nodes ADN, demonstrating that decomposition can significantly improve the speed of solving large-scale problems while maintaining accuracy with low AEV penetration.
\end{abstract}
\IEEEpeerreviewmaketitle
\renewcommand\nomgroup[1]{%
  \item[\bfseries
  \ifstrequal{#1}{A}{Sets}{%
  \ifstrequal{#1}{B}{Parameters}{%
  \ifstrequal{#1}{C}{Variables}{}}}%
]}

\nomenclature[A,01]{\(N\)}{Set of nodes}
\nomenclature[A,02]{\(L\)}{Set of distribution lines}
\nomenclature[A,03]{\(T\)}{Set of time intervals}
\nomenclature[A,04]{\(AEV\)}{Set of autopilot electric vehicles (AEVs)}
\nomenclature[A,05]{\(V2G\)}{Set of V2G charging stations (V2GCS)}
\nomenclature[A,06]{\(\Omega\)}{Set of areas}

\nomenclature[B,01]{\(\tau_u\)}{Time period AEV $u$ is connected to the power grid}
\nomenclature[B,02]{\(\varphi_{i}^{V2G}\)}{Retrofit or construction cost of V2GCS at node $i$}
\nomenclature[B,03]{\(\varphi_{ij}\)}{Construction cost of line $ij$}
\nomenclature[B,04]{\(R_{ij},X_{ij}\)}{Resistance and reactance of line $ij$}
\nomenclature[B,05]{\(C_{tou}\)}{Time of use price}
\nomenclature[B,06]{\(P_{i,t}^{Load},Q_{i,t}^{Load}\)}{Active / reactive power load at node $i$ in time $t$}
\nomenclature[B,08]{\(E_{u,0}^{AEV},E_{u}^{AEV}\)}{Initial and target energy of AEV $u$}
\nomenclature[B,09]{\(C_{ch,t}^{AEV},C_{dis,t}^{AEV}\)}{Charge and discharge price of AEV at time $t$}
\nomenclature[B,10]{\(\eta_{ch,i}^{ESS},\eta_{dis,i}^{ESS}\)}{Charge and discharge efficiency coefficient of the energy storage system (ESS) at node $i$}
\nomenclature[B,11]{\(\underline {E_u^{AEV}},\overline {E_u^{AEV}}\)}{Minimum and maximum energy capacity of AEV $u$ }
\nomenclature[B,12]{\(\underline {p_u^{AEV}},\overline {p_u^{AEV}}\)}{Maximum charging and discharging power of AEV $u$}
\nomenclature[B,13]{\(\underline {S_i^{V2G}},\overline {S_i^{V2G}}\)}{Minimum and maximum capacity of V2GCS at node $i$}
\nomenclature[B,14{\(\overline {P^{Sub}}\)}{Maximum active power of substation}
\nomenclature[B,15]{\(\overline {Q^{Sub}}\)}{Maximum reactive power of substation}
\nomenclature[B,16]{\(\underline {V},\overline {V}\)}{Minimum and maximum voltage magnitude}
\nomenclature[B,17]{\(\overline {P_{i,t}^{PV}}\)}{Maximum active power of photovoltaic (PV) at node $i$ in time $t$}
\nomenclature[B,18]{\(\underline {Q_{i}^{SVC}},\overline {Q_{i}^{SVC}}\)}{Minimum and maximum reactive power of static var compensator (SVC) at node $i$}
\nomenclature[B,19]{\(\underline {P_{i}^{ESS}},\overline {P_{i}^{ESS}}\)}{Minimum and maximum active power of the ESS at node $i$}
\nomenclature[B,20]{\(\underline {E_{i}^{ESS}},\overline {E_{i}^{ESS}}\)}{Minimum and maximum capacity of the ESS at node $i$}
\nomenclature[B,21]{\(Q_{min}^{CB}\)}{The minimal reactive power of capacitor bank (CB)}
\nomenclature[B,22]{\(v_{s}^{CB}\)}{Reactive power of increasing per bank of CB}
\nomenclature[B,23]{\(N^{CB}\)}{Maximum number of bank regulations for CB}
\nomenclature[B,24]{\(\overline {N_{i}^{CB}}\)}{Maximum banks of CB to be installed at node $i$}
\nomenclature[B,25]{\(V_{min}^{Oltc}\)}{The minimal voltage adjusted by on-load tap changer (OLTC)}
\nomenclature[B,26]{\(v_{s}^{Oltc}\)}{Voltage of increasing per tap step of OLTC}
\nomenclature[B,27]{\(N^{Oltc}\)}{Maximum number of step regulations for OLTC}
\nomenclature[B,28]{\(\overline {N^{Oltc}}\)}{Maximum variation of tap steps of the OLTC}
\nomenclature[B,29]{\(\overline {S_{ij}}\)}{Apparent power capacity for line $ij$}
\nomenclature[B,30]{\(\varphi_{i}^{DGRs}\)}{Construction cost of DGRs (including ESS, CB, PV, SVC) at node $i$}
\nomenclature[B,31]{\(c_{i}^{DGRs}\)}{Annualized operational maintenance cost of DGRs (including ESS, CB, PV, SVC) at node $i$}
\nomenclature[B,32]{\(c_{i}^{V2G}\)}{Annualized operational maintenance cost of V2GCS at node $i$}

\nomenclature[C,00]{\(\)}{}
\nomenclature[C,01]{\(p_{ch,u,t}^{AEV},p_{dis,u,t}^{AEV}\)}{Active and reactive power of AEV $u$ charging and discharging at time $t$}
\nomenclature[C,02]{\(P_{i,t}^{V2G},Q_{i,t}^{V2G}\)}{Active and reactive power of V2GCS at node $i$ in time $t$}
\nomenclature[C,03]{\(P_{t}^{Sub},Q_{t}^{Sub}\)}{Active / reactive power of substation in time $t$}
\nomenclature[C,05]{\(V_{i,t}\)}{Voltage at node $i$ in time $t$}
\nomenclature[C,06]{\(P_{ij,t},Q_{ij,t}\)}{Active and reactive power flow in line $ij$ at time $t$}
\nomenclature[C,07]{\(P_{i,t}^{PV}\)}{Active power of PV at node $i$ in time $t$}
\nomenclature[C,08]{\(Q_{i,t}^{SVC}\)}{Reactive power of SVC at node $i$ in time $t$}
\nomenclature[C,09]{\(E_{i,t}^{ESS}\)}{Energy storage of ESS at node $i$ in time $t$}
\nomenclature[C,10]{\(P_{ch,i,t}^{ESS},P_{dis,i,t}^{ESS}\)}{Charge and discharge power of ESS at node $i$ in time $t$}
\nomenclature[C,11]{\(T_{ch,i,t}^{ESS},T_{dis,i,t}^{ESS}\)}{Binary variable associated with charge and discharge status of ESS at node $i$ in time $t$}
\nomenclature[C,12]{\(Q_{i,t}^{CB}\)}{Reactive power of CB at node $i$ in time $t$}
\nomenclature[C,13]{\(T_{s,i,t}^{CB}\)}{Binary variable associated with bank quantity $s$ of CB at node $i$ in time $t$}
\nomenclature[C,14]{\(T_{in,i,t}^{CB},T_{de,i,t}^{CB}\)}{Binary variable associated with the increase and decrease status of CB at node $i$ in time $t$}
\nomenclature[C,15]{\(V_{i,t}^{Oltc}\)}{Voltage adjusted by OLTC at node $i$ in time $t$}
\nomenclature[C,16]{\(T_{s,i,t}^{Oltc}\)}{Binary variable associated with step quantity $s$ of OLTC at node $i$ in time $t$}
\nomenclature[C,17]{\(T_{in,i,t}^{Oltc},T_{de,i,t}^{Oltc}\)}{Binary variable associated with the increase and decrease status of OLTC at node $i$ in time $t$}
\nomenclature[C,18]{\(y_{i}^{V2G}/p_{u}/z_{ij}\)}{Binary variable associated with V2GCS at node $i$ / AEV $u$ / line $ij$}
\nomenclature[C,19]{\(y_{i}^{DGRs}\)}{Binary variable associated with construction of DGRs (including ESS, CB, PV, SVC) at node $i$}

\printnomenclature[2.2cm]
\section{Introduction}

\IEEEPARstart{W}{ith} the increasing concerns regarding global climate change and the depletion of fossil fuels, there has been a growing interest in incorporating distributed generation resources (DGRs) and electric vehicles (EVs) into distribution systems \cite{1}. The power sector is currently experiencing a major transformation in system planning, operation, and control paradigms aimed at achieving a secure and cost-effective energy transition \cite{2}. Additionally, advanced driver assistance technology has matured and is effectively being employed in the market \cite{33}.

Various DGRs have improved power quality and maintained stable operation of the distribution network. However, due to the diverse behavior of EVs, their charging demand growth is non-uniform in time and location, leading to disproportionate peaks \cite{3}. Furthermore, large-scale EV charging can cause power grid security issues such as significant power losses \cite{4}, voltage drops \cite{5}, and variations \cite{6}.

Nonetheless, EVs equipped with vehicle-to-grid (V2G) functionality hold the potential to facilitate the integration of DGRs \cite{35}. The intermittent nature of DGRs presents challenges to grid stability. While V2G technology empowers EVs to store excess renewable energy during periods of high generation and discharge it back to the grid during times of high demand. Therefore, this bidirectional energy flow of V2G enables EVs to serve as energy storage units, allowing grid operators to balance supply and demand effectively \cite{36}. Furthermore, EVs can be utilized as controllable resources, providing ancillary services to the distribution system, including peak shaving, voltage regulation, and stability enhancement \cite{7}. Thus, the optimal planning of networks that incorporate both V2G charging stations (V2GCS) and DGRs becomes a critical task in modern grid planning \cite{8}.

In the existing literature, various DGRs have been proposed for optimizing operations, siting, and capacity planning \cite{9}. For example, in \cite{11}, the authors proposed an energy storage system (ESS) planning model that considers investment costs, system maintenance costs, and deferred equipment investments. In \cite{12}, a mixed-integer linear programming (MILP) model was introduced to solve the problem of sizing and siting wind and solar power generating units in radial distribution systems. The objective of this approach is to minimize the system's investment and operating costs. Reference \cite{27} focused on the integrated planning problem of cyber-physical distribution networks and modeled multiple DGRs, including capacitor banks (CB) and energy storage systems (ESS). Moreover, \cite{10} and \cite{32} planned the siting and sizing of multiple DGRs, such as photovoltaic (PV) systems, static var compensation (SVC), on-load tap changer (OLTC), to minimize the annual operating costs of the distribution system.

In addition to DGRs, researchers have also studied the role of EVs in the distribution network. In \cite{13}, the uncertainty of EV behavior was considered, and the location and capacity of DGR investments were planned to reduce power losses caused by uncertainty. Reference \cite{14} used a three-stage optimization model with a gradually reduced time horizon to improve the voltage quality of the system by regulating the charging and discharging behavior of EVs. An MILP model was proposed in \cite{15} to optimize the charging behavior of EVs in unbalanced electrical distribution systems. Furthermore, reference \cite{16} introduced a multi-agent reinforcement learning algorithm based on the deep deterministic policy gradient method, considering the collaborative control problem of simultaneous active-reactive interaction between EVs and grid. However, these articles focused on improving EV performance in a fixed distribution network, rather than a dynamic one. Reference \cite{37} introduced a two-stage distributionally robust optimization model for the joint planning of EV charging stations and the distribution network. Additionally, references \cite{38} and \cite{39} addressed the planning of different charging facilities, focusing on integrated system and charging-battery swapping stations, respectively. While these studies extensively examined traffic flow and uncertainty, they offered limited insights into multiple DGRs and V2G functionality.

Reference \cite{17} proposed a coordinated optimal planning model for V2GCS and multiple DGRs, which considers multiple planning objectives, including system investment cost, reliability, power losses, and voltage stability. Reference \cite{18} presented a planning model for sizing and siting V2GCS, ESS, and other DGRs. However, this model was based on a fixed distribution network and did not account for future network expansion. To address this limitation, reference \cite{19} introduced a natural aggregation algorithm to plan the location and capacity of V2GCS, PV systems, and ESS, while considering the queuing time and minimizing investment network loss. In \cite{20}, a sequential capacitated flow capturing location allocation model was proposed for planning the distribution network and V2GCS. Additionally, reference \cite{21} developed a mixed integer second-order cone programming (MISOCP) model for solving multiple DGRs and location planning for V2GCS. Moreover, reference \cite{22} developed an MILP model that integrates the needs of both the traffic network and the distribution network using a network modeling approach based on a winner-takes-all edge trimming technique to identify interest points of the city in terms of traffic flows. Despite these efforts, the bi-directional interaction of active and reactive power when planning distribution networks and V2GCS has not been considered in the aforementioned studies.

This paper proposes a comprehensive model that facilitates the joint deployment of the ADN and V2GCS, while accounting for the bi-directional active-reactive power interaction of EVs. Our key contributions are outlined below:

1) We propose a novel collaborative planning model for the joint deployment of the ADN and V2GCS. This model takes into full consideration the V2G functionality of EVs and efficiently utilizes the unused capacities of V2G inverters to provide reactive power support for the grid.

2) A sequential decomposition method is proposed, transforming the holistic problem into two sub-problems, based on the weak coupling property of the physical problem. This approach not only increases the solving speed while ensuring the accuracy at low EV penetration, but also yields high-quality approximate solutions for intractable problems at high EV penetration.

3) This paper models multiple DGRs in the future ADN, where autonomous driving technology is extensively deployed. The control variables encompass active power, reactive power, and system voltage, with control strategies encompassing both continuous and discrete adjustments.

This paper is organized as follows. In Section II, we present the assumptions and simplifications in the model, as well as the existence analysis of solutions and rationale for the sequential decomposition. Section III describes the mathematical model for decomposition method. Section IV presents the illustration of cases and simulation results. Finally, in Section V, we provide the conclusion.

\section{Model Formulation}

\subsection{Assumptions and Simplifications in the Modeling}

To facilitate understanding and improve the efficiency of the proposed model, certain assumptions and simplifications were made during the modeling process. This subsection provides a detailed description of these assumptions and simplifications.

1) In the future transportation system, EV users will be directed by the information processing center, while the intelligent transportation system will guide users to charge their vehicles in designated areas.

2) In the context of the widespread adoption of Autopilot EV (AEV), these vehicles have the capability to arrive at the V2GCS charging station promptly at the user's scheduled time and ensure an ample battery level to meet the user's intended off-grid requirements.

3) The behavior of EVs after clustering provides a better representation of AEVs in different areas, presented in the appendix.

\subsection{Framework of the Sequential Decomposition Method}

The holistic collaborative planning model for ADN and V2GCS has been formulated, taking into account the behavior of AEVs. The model is a large-scale MISOCP problem, with the objective function and constraints detailed in equations \eqref{eq53} and \eqref{eq54}, respectively.

\centerline {Holistic Problem Formulation: ( large-scale MISOCP )}
\vspace{-1em}
\begin{equation}
\underset{x \in X, y \in Y}{min} \quad f(x)+E_{\xi}[q(y,\xi)]\label{eq53}
\end{equation}   
\vspace{-0.5em}
\begin{equation}
s.t.\left\{
\begin{aligned}
& G(\xi)y=d(\xi) \\
& N(\xi)y \geq b(\xi) \\
& T(\xi)x+W(\xi)y=h(\xi) \\
& F(\xi)x+H(\xi)y \geq u(\xi) \\
& E_{\xi}[q(y,\xi)]=\sum_{k=1}^{K}p_{k}q(y,\xi_{k})\label{eq54}
\end{aligned}
\right.
\end{equation}

In this formulation, the variables $x$ represent construction of V2GCS and planning of ADN, while the variables $y$ represent charging and discharging behavior of AEVs. The uncertain set of AEV behavior for stochastic programming is represented by $\xi(q,G,N,T,W,F,H,d,b,h,u)$, which includes $K$ scenarios with respective probability masses $p$. The expectation of the objective function in different scenarios is represented by $E$.

For large-scale MISOCP problems, the convergence speed may decrease, making the problem difficult to solve. To address this issue in the context of AEVs planning, this paper proposes a sequential decomposition method that decomposes the large-scale MISOCP problem into MILP and MISOCP problems, as shown in equations \eqref{eq55}-\eqref{eq58}.

\begin{figure*}[htbp]
    \centering
    \includegraphics[width=0.96\textwidth]{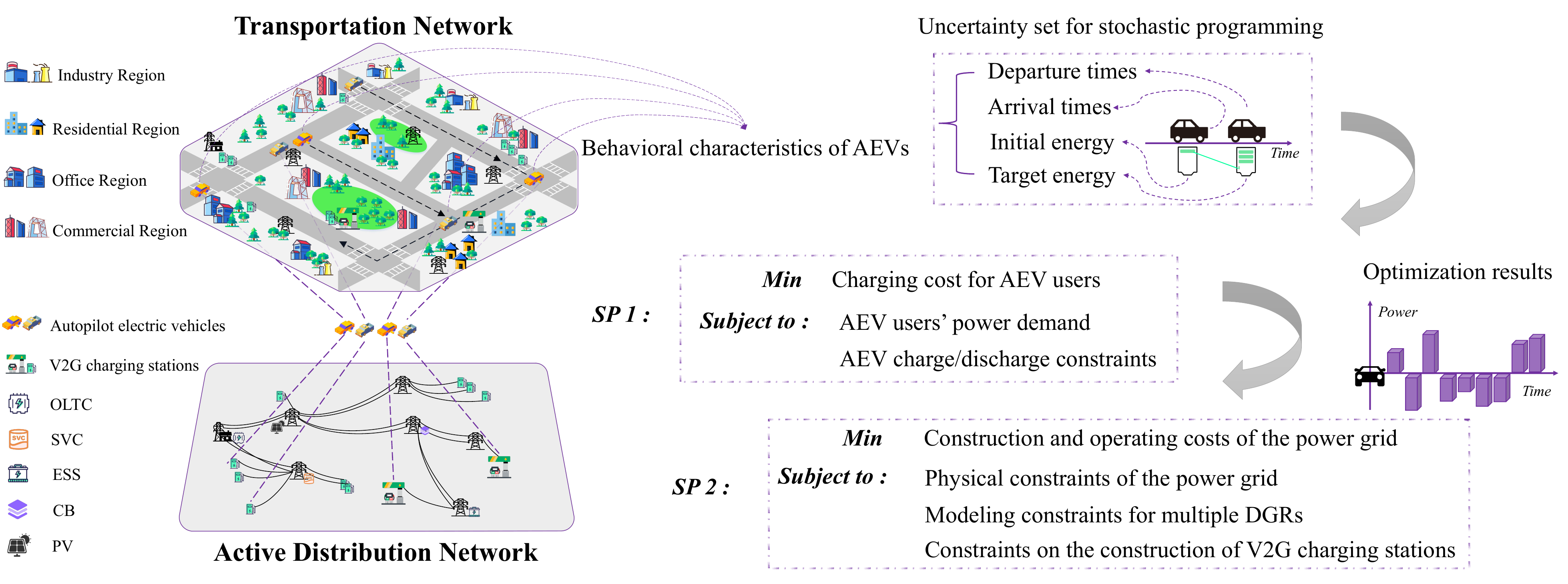}
    \caption{Optimization framework for sequential decomposition method}\label{fig1}
\end{figure*}

\centerline {Subproblems: ( MILP and MISOCP )}

\textit{Subproblem I (SP1}): ( MILP )
\vspace{-0.5em}
\begin{equation}
\underset{y \in Y}{min} \quad q^{T}(\xi) y\label{eq55}
\end{equation}   
\vspace{-0.5em}
\begin{equation}
s.t.\left\{
\begin{aligned}
& G(\xi) y=d(\xi) \\
& N(\xi)y \geq b(\xi)\label{eq56}
\end{aligned}
\right.
\end{equation}   

\textit{Subproblem II (SP2}): ( MISOCP )
\vspace{-0.5em}
\begin{equation}
\underset{x \in X}{min} \quad f(x)+E_{\xi}[Q(y^{*},\xi)]\label{eq57}
\end{equation}   
\vspace{-0.5em}
\begin{equation}
s.t.\left\{
\begin{aligned}
& T(\xi)x+W(\xi)y^{*}=h(\xi) \\
& F(\xi)x+H(\xi)y^{*} \geq u(\xi)\\
& E_{\xi}[Q(y^{*},\xi)]=\sum_{k=1}^{K}p_{k}Q(y^{*},\xi_{k})\label{eq58}
\end{aligned}
\right.
\end{equation}

In \textit{SP1}, the charging and discharging power of AEVs $y$ is optimized with the uncertainty set $\xi$ according to their behavioral characteristics. In \textit{SP2}, the distribution network and V2GCS are collaboratively planned based on the optimal value $Q(y,\xi)$ obtained in \textit{SP1}. Additionally, the sitting and sizing of DGRs are also planned. The physical model of the sequential decomposition method is shown in Fig. \ref{fig1}.

\subsection{Feasibility of the Sequential Decomposition Method}

In the context of employing the sequential decomposition method for problem-solving, an indispensable prerequisite for the presence of a solution entails that the outcome $y^{*} \in Y$, derived from \textit{SP1}, producing a feasible \textit{SP2}. Essentially, this implies the existence of an $x^{*} \in X$ that satisfies the subsequent equation (i.e., the constraints of \textit{SP2}):

\begin{equation}
\left\{
\begin{aligned}
& T(\xi)x^{*}+W(\xi)y^{*}=h(\xi) \\
& F(\xi)x^{*}+H(\xi)y^{*} \geq u(\xi) \label{eq59}
\end{aligned}
\right.
\end{equation}

We found out by repeated tests that, with the proposed sequential decomposition method, it's highly impossible that \textit{SP2} is infeasible with $y^{*}$. However, if someone concerns the feasibility of decomposition, a quick verification of solution existence by method could be conducted, as illustrated in Table \ref{Proof}.
In Step I, produce the "worst case" in \textit{SP1} by assuming that AEVs at V2GCS are always charging, neglecting the state of charge, denoted by $\overline{y^{*}}$. In Step II, incorporate the worst case $\overline{y^{*}}$ into the constraints of \textit{SP2}. If \textit{SP2} is feasible with $\overline{y^{*}}$, then the sequential decomposed model must be feasible. It should be noted that, the constraints \eqref{eq59} in \textit{SP2} denote the power flow constraints, security constraints, and V2GCS capacity constraints in ADN, as shown in Table \ref{Proof}.

 \begin{figure}[htbp]
    \centering
    \includegraphics[width=0.25\textwidth]{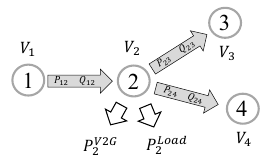}
    \caption{A small example for feasibility analysis}\label{fig11}
\end{figure}

Fig. \ref{fig11} provides a small example of ADN for discussion. If the V2GCS is located at node 2, the power injection $P_{2}^{V2G}$ influences the power flows, e.g. $P_{12}, P_{23}, P_{24}$. Nevertheless, compared to other types of loads (e.g., air conditioning in urban areas), $P_{2}^{V2G}$ is generally smaller, and its impact on other power flows usually remains within the distribution line capacity. The scenario where the AEV connected to the system is simultaneously charging at its maximum power represents the highest load profile for $P_{2}^{V2G}$, denoted by $\overline{y_{2}^{V2G}}$. If $SP2$ remains feasible with $\overline{y_{2}^{V2G}}$, then the feasibility of the sequential decomposition method is guaranteed with arbitrary AEVs charging behavior.
 
\begin{table}[htbp]
\caption{Verification of Solution Existence\\(based on the Example in Fig. 2)}
\label{Proof}
\scriptsize
\begin{tabular}{ll}
\hline \hline
\textit{\textbf{Step I:}} 
& Produce the "worst case" in \textit{SP1} by assuming that AEVs\\
& at V2GCS are always charging, denoted by $\overline{y_{2}^{V2G}}$.\\ \hline
\textit{\textbf{Step II:}} 
& Incorporating the worst case $\overline{y_{2}^{V2G}}$ into the constraints of\\
& \textit{SP2}, as follows: \\ \hline
\textit{Power Balance:} 
&$P_{12}-P_{23}-P_{24}=\overline{y_{2}^{V2G}}+P_{2}^{Load}$\\
&$Q_{12}-Q_{23}-Q_{24}=Q_{2}^{Load}$\\
\textit{Security Constraints:}
&$\underline{V} \le V_{1},V_{2},V_{3},V_{4} \le \overline{V}$\\
&$\underline{S} \le P_{12}^{2}+Q_{12}^{2} \le \overline{S}$\\
&$\underline{S} \le P_{23}^{2}+Q_{23}^{2} \le \overline{S}$\\
&$\underline{S} \le P_{24}^{2}+Q_{24}^{2} \le \overline{S}$\\
\textit{Capacity Constraints:}
&$\overline{y_{2}^{V2G}} \le S_{2}^{V2G}$ \\ \hline
\textit{\textbf{If satisfied,}} 
& There must be a solution for the decomposed method \\
\textit{\textbf{If not,}} 
& Solutions do not necessarily exist (extremely rare events) \\
\hline \hline
\end{tabular}
\end{table}

\subsection{Rationale for Sequential Decomposition}

The possibility of a feasible decomposition lies in the weak coupling between variables $x$ and $y$, as shown in constraints of equation \eqref{eq54}. Since EVs possess autopilot and V2G characteristics, the position of V2GCS ($x$) has a minimal impact on the EV charging and discharging power ($y$). Conversely, the $y$ only affects the power flow distribution of the distribution network, as reflected in the objective function \eqref{eq53}, in other words, $y$ only impacts the network loss: a small proportion of the total objective function. Therefore, the coupling relationship between $x$ and $y$ is weak, allowing them to be separated into two subproblems for solving.

During peak load periods, time-of-use (TOU) pricing is higher, which optimizes AEV behavior in subproblem I based on price. This is equivalent to considering some of the grid demand.
Additionally, the large number of AEVs and their diverse behavioral characteristics make it possible to obtain almost identical overall load of the V2GCS after the AEV load superposition.

\section{Mathematical Model}

\subsection{SP 1: Optimization of AEV charging behavior}
\begin{equation}
min \sum_{\substack{t \in \tau_u}}\sum_{\substack{u \in AEV}}C_{ch,t}^{AEV}p_{ch,u,t}+\sum_{\substack{t \in \tau_u}}\sum_{\substack{u \in AEV}}C_{dis,t}^{AEV}p_{dis,u,t} \label{eq1}  
\end{equation}
Subject to: \eqref{eq2}-\eqref{eq9}

The objective function \eqref{eq1} minimizes the AEV charging costs. The subscript $ch$ and $dis$ denote the charging and discharging states, respectively, while $t$ indicates the period during which AEV $u$ is connected to the power grid. AEV charging incurs a service charge $ \rm C_{ch}^{AEV}$ based on a TOU tariff, while a battery compensation price incentive $ \rm C_{dis}^{AEV}$ is set to encourage AEVs to deliver active power to the grid without compromising their behavioral characteristics.

\begin{flalign*}
&\underline{E_{u}^{AEV}} \le E_{u,0}^{AEV}(\xi)+\sum_{\substack{t}}p_{ch,u,t}^{AEV}+\sum_{\substack{t}}p_{dis,u,t}^{AEV}\le \overline{E_{u}^{AEV}} &  \label{eq2}
\end{flalign*}
\vspace{-1.5em}
\begin{flalign}
&&\forall u \in AEV, \forall t \in \tau_u
\end{flalign}
\vspace{-1.7em}
\begin{flalign*}
& E_{u,0}^{AEV}(\xi)+\sum_{\substack{t \in \tau_u}}p_{ch,u,t}^{AEV}+\sum_{\substack{t \in \tau_u}}p_{dis,u,t}^{AEV} \geq E_{u}^{AEV}(\xi) &\label{eq3}
\end{flalign*}
\vspace{-2em}
\begin{flalign}
&&\forall u \in AEV
\end{flalign}
\vspace{-2em}
\begin{flalign}
&\sum_{\substack{t \in T-\tau_u}}\sum_{\substack{u \in AEV}}p_{ch,u,t}^{AEV}+\sum_{\substack{t \in T-\tau_u}}\sum_{\substack{u \in AEV}}p_{dis,u,t}^{AEV}=0 & \label{eq4}  
\end{flalign}
\vspace{-1em}
\begin{flalign}
& p_{ch,u,t}^{AEV} \le (1-p_{u}) M &\label{eq5}
& & \forall u \in AEV, \forall t \in \tau_u &
\end{flalign}
\vspace{-2em}
\begin{flalign}
& p_{dis,u,t}^{AEV} \ge -p_{u} M &\label{eq6}
& & \forall u \in AEV, \forall t \in \tau_u &
\end{flalign}
\vspace{-2em}
\begin{flalign}
& 0 \le p_{ch,u,t}^{AEV} \le \overline{p_{u}^{AEV}} &\label{eq7}
& & \forall u \in AEV, \forall t \in \tau_u &
\end{flalign}
\vspace{-2em}
\begin{flalign}
& \underline{p_{u}^{AEV}} \le p_{dis,u,t}^{AEV} \le 0 &\label{eq8}
& & \forall u \in AEV, \forall t \in \tau_u &
\end{flalign}
\vspace{-2em}
\begin{flalign}
& p_{u} \in \{0,1\} &\label{eq9}
& & \forall u \in AEV &
\end{flalign}

The capacity constraints of AEVs are expressed in equation \eqref{eq2}, which limits the energy of each AEV within appropriate ranges. Equation \eqref{eq3} represents the battery capacity required to meet the target energy ($E_{u}^{AEV}(\xi)$) when the AEV departs from the power grid. Equation \eqref{eq4} constrains the power of AEVs to zero before arriving and after departure from the V2GCS. Constraints \eqref{eq5} to \eqref{eq8} define the charging and discharging power constraints using the big $M$ method. The constant M is chosen to be large enough to relax the inequalities \eqref{eq5} and \eqref{eq6}. If AEV $u$ is charging in time $t$, i.e. $p_{u}$=0, the corresponding constraint will be enforced and the AEV cannot discharge in the time $t$. Equations \eqref{eq7} and \eqref{eq8} indicate that the charging and discharging power ($p_{ch,u,t}^{AEV}$,$p_{dis,u,t}^{AEV}$) should be limited between the upper and lower bounds ($\underline {p_u^{AEV}}$,$\overline {p_u^{AEV}}$).

\subsection{SP 2: Coordinated Planning of ADN and V2GCS}
\vspace{-1.2em}
\begin{flalign}
min \quad C_{line}^{Inv}+C_{V2G}^{Inv}+C_{DGRs}^{Inv}+C_{ADN}^{O\&M}+C_{loss}^{Ope} \label{eq10}  
\end{flalign}
Subject to: \eqref{eq17}-\eqref{eq50} 
\vspace{-0.1em}
\begin{flalign}
C_{line}^{Inv}=R_{d}\sum_{\substack{ij \in L}}\varphi_{ij}z_{ij} \label{eq11}  
\end{flalign}
\vspace{-1em}
\begin{flalign}
C_{V2G}^{Inv}=R_{d}\sum_{\substack{i \in N}}\varphi_{i}^{V2G}y_i \label{eq12}  
\end{flalign}
\vspace{-1em}
\begin{flalign}
C_{DGRs}^{Inv}=R_{d}\sum_{\substack{i \in N}}\varphi_{i}^{DGRs}y_{i}^{DGRs} \label{eq13}  
\end{flalign}
\vspace{-1em}
\begin{flalign}
C_{ADN}^{O\&M}=\sum_{\substack{i \in N}}c_{i}^{V2G}y_i^{V2G}+\sum_{\substack{i \in N}}c_{i}^{DGRs}y_i^{DGRs} \label{eq14}  
\end{flalign}
\vspace{-1em}
\begin{flalign}
C_{loss}^{Ope}=365 \cdot \sum_{\substack{t \in T}}\sum_{\substack{ij \in L}}C_{t}^{TOU}  R_{ij}(P_{ij,t}^{2}+Q_{ij,t}^{2}) \label{eq15}  
\end{flalign}
\vspace{-1em}
\begin{flalign}
R_{d}=\frac{d(1+d)^{year}}{(1+d)^{year}-1} \label{eq16}  
\end{flalign}

The objective function \eqref{eq10} minimize the annualized investment cost, operational and maintenance expenses, and power system network loss. The subscript $ij$ denotes the $ij$-th line, $i$ denotes the $i$-th node, and $t$ represents the time interval.

The objective functions \eqref{eq11}-\eqref{eq13} represent the costs associated with the construction of distribution lines and DGRs (including PV, ESS, SVC, and CB), as well as the expenses of retrofitting or building V2GCS. Within these functions, $R_d$ represents the annualized cost coefficient, $d$ represents the inflation rate, and $year$ represents the operational lifespan. Objective function \eqref{eq14} minimizes the annualized operational and maintenance expenses of ADN, accounting for both the V2GCS and multiple DGRs, while objective function \eqref{eq15} minimizes the annualized network loss of the distribution grid. The mathematical model includes multiple DGRs, considering their regulation functions and operation modes. Related constraints are shown in Table \ref{tab:management}.

\begin{table}[ht]
  \centering
  \caption{Mathematical Model for Multiple DGRs}
  \label{tab:management}
  \begin{tabular}{cccc}
    \hline\hline
    Title    & Regulation Function   & Operation Mode   & Constraints  \\ \hline
    PV       & active power       & continuous & equation \eqref{eq30}         \\
    ESS      & active power       & continuous & equations \eqref{eq32}-\eqref{eq36}   \\
    CB       & reactive power     & discrete & equations \eqref{eq37}-\eqref{eq42}   \\ 
    SVC      & reactive power     & continuous & equation \eqref{eq31}       \\ 
    OLTC     & voltage & discrete   & equations \eqref{eq43}-\eqref{eq50} \\
    \hline\hline
  \end{tabular}
\end{table}

\noindent (1) \textit{Power Flow Constraints:}
\begin{flalign*}
& P_{\sim i,t}+P_{i,t}^{V2G}+y_{i}^{PV}P_{i,t}^{PV}+y_{i}^{ESS}(P_{ch,i,t}^{ESS}-P_{dis,i,t}^{ESS})& \label{eq17}
\end{flalign*}
\vspace{-2em}
\begin{flalign}
&& =P_{i\sim,t}+P_{i,t}^{Load} \qquad \qquad \forall ij \in L, \forall t \in T, \forall i \in N
\end{flalign}
\begin{flalign*}
& Q_{\sim i,t}+Q_{i,t}^{V2G}+y_{i}^{SVC}Q_{i,t}^{SVC}+y_{i}^{CB}Q_{i,t}^{CB}=Q_{i\sim,t}+Q_{i,t}^{Load} & \label{eq18}
\end{flalign*}
\vspace{-2em}
\begin{flalign}
&& \forall ij \in L, \forall t \in T, \forall i \in N
\end{flalign}
\vspace{-1.5em}
\begin{flalign*}
& \sqrt{V_{i,t}^2-V_{j,t}^2-2(R_{ij}P_{ij,t}+X_{ij}Q_{ij,t})} & \label{eq19}
\end{flalign*}
\vspace{-2em}
\begin{flalign}
&&  \le (1-z_{ij}) \times M \qquad  \forall ij \in L, \forall t \in T, \forall i,j \in N
\end{flalign}

The distflow model is widely adopted to describe the power flow in radial distribution networks \cite{23} \cite{28}, as shown in equations \eqref{eq17}-\eqref{eq19}. Equation \eqref{eq17} represents the active power balance at node $i$. The active power injected to the node includes the active power input of the lines connected to it ($P_{\sim i,t}$), the active power injection by the V2GCS ($P_{i,t}^{V2G}$), the PV contribution ($P_{i,t}^{PV}$), and the energy generated by ESS ($P_{ch,i,t}^{ESS}$). The active power output at node $i$ consists of the active power output of the line connected from it ($P_{i\sim,t}$), the active load at the node ($P_{i,t}^{Load}$), and energy absorbed by ESS ($P_{dis,i,t}^{ESS}$) \cite{24}. Similarly, equation \eqref{eq18} represents the balance of reactive power, comprising reactive power inflow through the line ($Q_{\sim i,t}$), the reactive power from the V2GCS ($Q_{i,t}^{V2G}$), the power contribution from SVC and CB ($Q_{i,t}^{SVC}$,$Q_{i,t}^{CB}$), the reactive power outflow through the line ($Q_{i\sim,t}$) and the reactive load ($Q_{i,t}^{Load}$). In equation \eqref{eq19}, $M$ is used to relax the inequality while $z_{ij}$=0. If the circuit is utilized in this scenario, i.e. $z_{ij}$=1, the corresponding constraint will be enforced \cite{26}.

\noindent (2) \textit{Radiality Constraints:}

Radiality constraints are involved in the distribution system, including spanning tree constraints and single-commodity flow-based radiality constraints \cite{29}.

\noindent (3) \textit{Security Constraints:}
\begin{flalign}
& P_{ij,t}^2+Q_{ij,t}^2 \le z_{ij} \times \overline{S_{ij}}^2 &\label{eq20}
& & \forall ij \in L, \forall t \in T &
\end{flalign}
\vspace{-2em}
\begin{flalign}
& \underline{V} \le V_{i,t} \le \overline{V} &\label{eq21}
& & \forall i \in N, \forall t \in T &
\end{flalign}

The line capacity and voltage magnitude are constrained by equations \eqref{eq20} and \eqref{eq21} to ensure system security.

\noindent (4) \textit{Substation Power Constraints:}
\begin{flalign}
& P_{i,t}^{Sub} \le \overline{P^{Sub}} &\label{eq22}
& & \forall t \in T &
\end{flalign}
\vspace{-2em}
\begin{flalign}
& Q_{i,t}^{Sub} \le \overline{Q^{Sub}} &\label{eq23}
& & \forall t \in T &
\end{flalign}

The active and reactive power flowing through a substation are restricted by the substation's capacity, as represented in equations \eqref{eq22} and \eqref{eq23}.

\noindent (5) \textit{V2GCS construction Constraints:}
\begin{flalign}
& -y_i \times M \le P_{i,t}^{V2G} \le y_i \times M &\label{eq24}
& & \forall i \in V2G, \forall t \in T &
\end{flalign}
\vspace{-1.5em}
\begin{flalign}
& -y_i \times M \le Q_{i,t}^{V2G} \le y_i \times M &\label{eq25}
& & \forall i \in V2G, \forall t \in T &
\end{flalign}
\vspace{-1.5em}
\begin{flalign*}
& -(1-y_i) \times M \le P_{i,t}^{V2G}-P_{i,t} \le (1-y_i) \times M &\label{eq26}
\end{flalign*}
\vspace{-2em}
\begin{flalign}
&& \forall i \in V2G, \forall t \in T
\end{flalign}
\vspace{-2em}
\begin{flalign*}
& -(1-y_i) \times M \le Q_{i,t}^{V2G}-Q_{i,t} \le (1-y_i) \times M &\label{eq27}
\end{flalign*}
\vspace{-2em}
\begin{flalign}
&& \forall i \in V2G, \forall t \in T
\end{flalign}
\vspace{-2em}
\begin{flalign*}
& \underline{S_i^{V2G}} \le (P_{i,t}^{V2G})^2+(Q_{i,t}^{V2G})^2 \le \overline{S_i^{V2G}} &\label{eq28}
\end{flalign*}
\vspace{-2em}
\begin{flalign}
&& \forall i \in V2G, \forall t \in T
\end{flalign}
\vspace{-2em}
\begin{flalign*}
& \sum_{\substack{i \in V2G}}P_{i,t}^{V2G}= \sum_{\substack{u \in AEV}}(p_{ch,u,t}^{AEV}+p_{dis,u,t}^{AEV})& \label{eq29}
\end{flalign*}
\vspace{-2em}
\begin{flalign}
&& \forall i \in \Omega, \forall u \in AEV, \forall t \in T
\end{flalign}

Equations \eqref{eq24}-\eqref{eq27} restrict the active and reactive power flowing through a V2GCS. The station's capacity ($P_{i,t}^{V2G}$,$Q_{i,t}^{V2G}$) and EVs' power ($p_{ch,u,t}^{AEV}$,$p_{dis,u,t}^{AEV}$) are constrained by \eqref{eq28} and \eqref{eq29}. If V2GCS is utilized in node $i$, i.e. $y_{i}$=1, constraints \eqref{eq24}-\eqref{eq27} will be relaxed, and the active and reactive power will be limited by capacity constraint \eqref{eq28}. Otherwise, if $y_{i}$=0, the V2GCS in node $i$ is neither retrofitted nor constructed. Constraint \eqref{eq29} means V2GCS are required to meet the charging demands of EVs at any $t$ and in any $\Omega$ (residential region, commercial region, industry region and office region). AEVs in the region with charging needs follow the scheduling instructions from the intelligent transportation system to charge at the corresponding V2GCS within the area.

\noindent (6) \textit{PV Operation Constraint:}
\begin{flalign}
    & 0 \le P_{i,t}^{PV} \le \overline {P_{i,t}^{PV}}&\label{eq30}
    & & \forall i \in N, \forall t \in T &
\end{flalign}
\noindent (7) \textit{SVC Operation Constraint:}
\begin{flalign}
    &\underline {Q_{i}^{SVC}} \le Q_{i,t}^{SVC} \le \overline {Q_{i}^{SVC}}& \label{eq31}
    & & \forall i \in N, \forall t \in T &
\end{flalign}
\noindent (8) \textit{ESS Operation Constraints:}
\begin{flalign*}
& E_{i,t}^{ESS}=E_{i,t-1}^{ESS}+P_{ch,i,t-1}^{ESS}\times \eta_{ch,i}^{ESS}+P_{dis,i,t-1}^{ESS}/\eta_{dis,i}^{ESS}&\label{eq32}
\end{flalign*}
\vspace{-2em}
\begin{flalign}
&&\forall i \in N, \forall t \in T
\end{flalign}
\vspace{-2em}
\begin{flalign*}
&\underline {P_{i}^{ESS}}\times{T_{ch,i,t}^{ESS}}\le{P_{ch,i,t}^{ESS}}\le \overline {P_{i}^{ESS}}\times{T_{ch,i,t}^{ESS}}&\label{eq33}
\end{flalign*}
\vspace{-2em}
\begin{flalign}
&&\forall i \in N, \forall t \in T
\end{flalign}
\vspace{-2em}
\begin{flalign*}
&\underline {P_{i}^{ESS}}\times{T_{dis,i,t}^{ESS}}\le{P_{dis,i,t}^{ESS}}\le \overline {P_{i}^{ESS}}\times{T_{dis,i,t}^{ESS}}&\label{eq34}
\end{flalign*}
\vspace{-2em}
\begin{flalign}
&&\forall i \in N, \forall t \in T
\end{flalign}
\vspace{-2em}
\begin{flalign}
& \underline {E_{i}^{ESS}}\le{E_{i,t}^{ESS}}\le\overline {E_{i}^{ESS}} & \label{eq35}
& &\forall i \in N, \forall t \in T &
\end{flalign}
\vspace{-2em}
\begin{flalign}
    & T_{ch,i,t}^{ESS}+T_{dis,i,t}^{ESS} \le 1 & \label{eq36}
    & & \forall i \in N, \forall t \in T &
\end{flalign}

The ESS capacity constraint is represented by expression \eqref{eq32}. The energy stored in the ESS at the beginning and end of an operation period must be equal. The charging power $P_{ch,i,t}^{ESS}$ and discharging power $P_{dis,i,t}^{ESS}$ must be limited within appropriate ranges as shown in \eqref{eq33} and \eqref{eq34}. It can be observed from the equations that the charging and discharging power range of the ESS is not only determined by the minimum and maximum ($ \rm \underline {P_{i}^{ESS}}$,$ \rm \overline {P_{i}^{ESS}}$), but also by the scheduling decision variables ($T_{ch,i,t}^{ESS}$,$T_{dis,i,t}^{ESS}$). Equation \eqref{eq35} dictates that the ESS's capacity should be limited between the upper and lower bounds. Expression \eqref{eq36} denotes the status change constraint of charging and the limitation of ESS operation.

\noindent (9) \textit{CB Operation Constraints:}
\begin{flalign*}
&Q_{i,t}^{CB} = Q_{min}^{CB} + \sum_{\substack{s}} v_{s}^{CB} \times T_{s,i,t}^{CB}&\label{eq37}
\end{flalign*}
\vspace{-2em}
\begin{flalign}
&&\forall i \in N, \forall t \in T, \forall s \in \overline{N_i^{CB}}
\end{flalign}
\vspace{-1.8em}
\begin{flalign*}
&\sum_{\substack{s}}T_{s,i,t}^{CB}-\sum_{\substack{s}}T_{s,i,t-1}^{CB}\le T_{in,i,t-1}^{CB} \times \overline{N_i^{CB}}-T_{de,i,t-1}^{CB}&\label{eq38}
\end{flalign*}
\vspace{-1.7em}
\begin{flalign}
&&\forall i \in N, \forall t \in T, \forall s \in \overline{N_i^{CB}}
\end{flalign}
\vspace{-1.8em}
\begin{flalign*}
&\sum_{\substack{s}}T_{s,i,t}^{CB}-\sum_{\substack{s}}T_{s,i,t-1}^{CB}\geq T_{in,i,t-1}^{CB}-T_{de,i,t-1}^{CB}\times \overline{N_i^{CB}}&\label{eq39}
\end{flalign*}
\vspace{-1.7em}
\begin{flalign}
&&\forall i \in N, \forall t \in T, \forall s \in \overline{N_i^{CB}}
\end{flalign}
\vspace{-1.8em}
\begin{flalign}
&\sum_{\substack{t}}T_{in,i,t}^{CB}+\sum_{\substack{t}}T_{de,i,t}^{CB}\le N^{CB} &\label{eq40}  
&& \forall i \in N, \forall t \in T &
\end{flalign}
\vspace{-1em}
\begin{flalign}
& T_{s,i,t}^{CB} \le T_{s-1,i,t}^{CB} & \label{eq41} 
& &\forall i \in N, \forall t \in T, \forall s \in \overline{N_i^{CB}} &
\end{flalign}
\vspace{-1.5em}
\begin{flalign}
& T_{in,i,t}^{CB} + T_{de,i,t}^{CB} \le 1 & \label{eq42} 
& &\forall i \in N, \forall t \in T &
\end{flalign}

The bank quantity of CB is denoted by $s$. For example, if the bank is set to the $5$-th position, the lower five binary variables are set to one, and the remaining binary variables above are set to zero. Expression \eqref{eq37} represents the discrete reactive power constraint of CB. Equations \eqref{eq38} and \eqref{eq39} restrict the regulation range of CB. It is evident that the regulation of CB is determined not only by the bank numbers but also by the scheduling decision variables ($T_{in,i,t}^{CB}$,$T_{de,i,t}^{CB}$). Expressions \eqref{eq41} and \eqref{eq42} represent the operation bank limits of CB and the status transition limit between on and off.

\noindent (10) \textit{OLTC Operation Constraints:}
\begin{flalign}
& V_{i,t}^{Sub}=V_{i,t}^{Oltc} &\label{eq43}
& & \forall i \in N, \forall t \in T &
\end{flalign}
\vspace{-1.8em}
\begin{flalign}
& \underline{V} \le V_{i,t}^{Oltc} \le \overline{V} &\label{eq44}  
&& \forall i \in N, \forall t \in T &
\end{flalign}
\vspace{-1.8em}
\begin{flalign*}
&V_{i,t}^{Oltc} = V_{min}^{Oltc} + \sum_{\substack{s}} v_{s}^{Oltc} \times T_{s,i,t}^{Oltc}&\label{eq45}
\end{flalign*}
\vspace{-2em}
\begin{flalign}
&&\forall i \in N, \forall t \in T, \forall s \in \overline{N_i^{Oltc}}
\end{flalign}
\vspace{-2em}
\begin{flalign*}
&\sum_{\substack{s}}T_{s,i,t}^{Oltc}-\sum_{\substack{s}}T_{s,i,t-1}^{Oltc}\le T_{in,i,t-1}^{Oltc} \times \overline{N_i^{Oltc}}-T_{de,i,t-1}^{Oltc}&\label{eq46}
\end{flalign*}
\vspace{-1.7em}
\begin{flalign}
&&\forall i \in N, \forall t \in T, \forall s \in \overline{N_i^{Oltc}}
\end{flalign}
\vspace{-2em}
\begin{flalign*}
&\sum_{\substack{s}}T_{s,i,t}^{Oltc}-\sum_{\substack{s}}T_{s,i,t-1}^{Oltc}\geq T_{in,i,t-1}^{Oltc}-T_{de,i,t-1}^{Oltc}\times \overline{N_i^{Oltc}}& \label{eq47}
\end{flalign*}
\vspace{-1.7em}
\begin{flalign}
&&\forall i \in N, \forall t \in T, \forall s \in \overline{N_i^{Oltc}}
\end{flalign}
\vspace{-2em}
\begin{flalign}
&\sum_{\substack{t}}T_{in,i,t}^{Oltc}+\sum_{\substack{t}}T_{de,i,t}^{Oltc}\le N^{Oltc} &\label{eq48}  
&& \forall i \in N, \forall t \in T &
\end{flalign}
\vspace{-1.2em}
\begin{flalign}
& T_{s-1,i,t}^{Oltc} \le T_{s,i,t}^{Oltc} & \label{eq49} 
& &\forall i \in N, \forall t \in T, \forall s \in \overline{N_i^{Oltc}} &
\end{flalign}
\vspace{-2em}
\begin{flalign}
& T_{in,i,t}^{Oltc} + T_{de,i,t}^{Oltc} \le 1 & \label{eq50} 
& &\forall i \in N, \forall t \in T &
\end{flalign} 

Equations \eqref{eq43} and \eqref{eq44} describe the utilization of the OLTC in the substation node, with the voltage of the OLTC constrained to ensure system security. Constraint \eqref{eq45} limits the voltage magnitude between upper and lower bounds. Equations \eqref{eq46} and \eqref{eq47} restrict the regulation range of the OLTC. The regulation of the OLTC is determined by the step numbers $ \rm \overline{N_i^{Oltc}}$ and scheduling decision variables ($T_{in,i,t}^{OLTC}$,$T_{de,i,t}^{OLTC}$). Expressions \eqref{eq49} and \eqref{eq50} represent the operation limits of the OLTC, and the constraint of status.

\section{Case Studies}

The proposed model has been tested in both the 47 nodes region of Shenzhen, China and the IEEE 33 nodes distribution network. Within the 47 nodes region, nodes 1-11 represent office areas, 12-17 are designated for industrial purposes, while nodes 18-33 and 34-47 are residential and commercial areas, respectively. In the residential region, a total of 144 AEVs are dispersed, whereas within the office region, the number of AEVs distributed amounts to 574, and in the industrial region, there is a distribution of 262 AEVs.

To ensure that voltage magnitude remains within acceptable limits, the maximum and minimum values were set at 1.1 pu and 0.9 pu, respectively. For Tesla Model S EVs, the charging and discharging power limits are set at 12 kW, with an energy capacity of 90 kWh. The substation node 1 to the bulk network is equipped with OLTC, while the maximum contribution of PV typically occurs at 14:00, reaching 75 kW. The upper and lower regulation limits for the SVC are set at 250 kvar and -50 kvar. Additionally, ESS was given an energy capacity of 800 kWh and maximum charging and discharging power limits of 200 kW and 300 kW, respectively, with the efficiency factors set at 0.9 and 1.1. CB has a maximum value of 375 kvar and can be regulated up to five times per day. OLTC, which has 20 steps ranging from 0.9 to 1.1, regulates up to six times per day. Please refer to Table \ref{tab:construction costs} for information about the construction and O\&M costs. Considering the future development of the region, it is anticipated that the number of EVs may further increase. Therefore, we consider the cost of relatively larger-scale stations to meet the demands of users.

\begin{table}[htbp]
  \centering
  \caption{Construction and O\&M Costs ($10^{4}$ CNY \textyen)}
  \label{tab:construction costs}
  \tabcolsep=3pt
  \begin{tabular}{ccccc}
    \hline \hline
    \thead{Title}      & \thead{Mode}  & \thead{Candidate Nodes} & \thead{Cost} & \thead{Economic life} \\ \hline
    \multirow{3}{*}{V2GCS}  & \multirowcell{1}{O\&M} & 1  --  47 & 4.70 & 1  \\
                            & \multirowcell{1}{Retrofit} & 17, 26, 27, 33, 47 & 84.97     & 10  \\
                            & \multicolumn{1}{c}{\makecell{Construction}} & 1-16,18-25,28-32,34-46 & 194.36 & 10 \\ \hline
    \multirow{2}{*}{PV}   & \multirowcell{1}{O\&M} & 1  --  47 & 0.50 & 1  \\
                            & \multicolumn{1}{c}{\makecell{Construction}} & 1  --  47 & 17.65 & 15 \\ \hline
    \multirow{2}{*}{ESS}   & \multirowcell{1}{O\&M} & 1  --  47 & 1.34 & 1  \\
                            & \multicolumn{1}{c}{\makecell{Construction}} & 1  --  47 & 24.94 & 20 \\ \hline
    \multirow{2}{*}{CB}   & \multirowcell{1}{O\&M} & 1  --  47 & 0.55     & 1  \\
                            & \multicolumn{1}{c}{\makecell{Construction}} & 1  --  47 & 10.38 & 15 \\ \hline
    \multirow{2}{*}{SVC}   & \multirowcell{1}{O\&M} & 1  --  47 & 0.65     & 1  \\
                            & \multicolumn{1}{c}{\makecell{Construction}} & 1  --  47 & 11.85 & 20 \\ \hline
    \multirow{1}{*}{Line}   & \multicolumn{1}{c}{\makecell{Construction}} & --- & 23.30 / km & 20 \\ \hline \hline
  \end{tabular}
\end{table}

\begin{figure*}[htbp]
    \centering
    \includegraphics[width=1.00\textwidth]{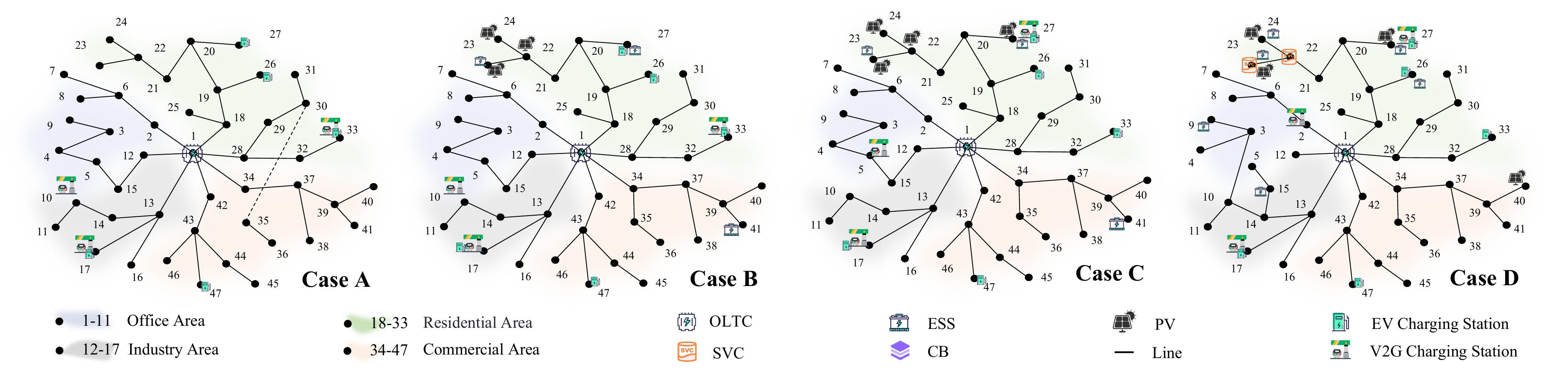}
    \caption{Planning solutions for different cases in Longgang District, Shenzhen, China}\label{fig3}
\end{figure*}

The TOU tariff is as shown in equation (58). When AEVs are charging, they will incur the same charging cost as the TOU. Conversely, when discharging, they will receive a subsidy equal to the TOU.

\begin{gather}
C_{t}^{TOU} = 
\begin{cases}
1.1121 \quad t \in [9,15) \cup [19,22) \notag \\
0.6542 \quad t \in [7,9) \cup [15,19) \cup [22,24) \\
0.2486 \quad t \in [1,7) \cup \{24\}
\end{cases} \notag \\
( Unit: yuan / kWh )
\end{gather}

Similar to DGRs, the AEV penetration rate in this paper denotes the ratio of the total capacity of AEVs involved in interaction to the overall power system load, as demonstrated in equation \eqref{eq52} \cite{19}.

\begin{flalign}
\qquad AEV \quad penetration \quad rate = \frac{\sum_{\substack{u}}\sum_{\substack{t}}\overline{E_{u,t}^{EV}}}{\sum_{\substack{i}}\sum_{\substack{t}}P_{i,t}^{Load}} \qquad \label{eq52}  
\end{flalign}

\subsection{Simulation Results}

To showcase the functionality of multiple DGRs, the advantages of bi-directional active-reactive power interaction of V2GCS, and the effectiveness of the proposed method in the paper, four distinct cases are presented. The model was formulated using the YALMIP tool in MATLAB (2021B) and evaluated with the GUROBI Optimizer (9.5.2) on the M1Pro chip, which boasts an 8-core CPU and a 14-core GPU \cite{25}. The four cases are outlined as follows:

\textit{Case A:} Traditional distribution network planning considering bi-directional active-reactive power interaction using sequential decomposition method

\textit{Case B:} Active distribution network planning considering bi-directional active power interaction using sequential decomposition method

\textit{Case C:} Active distribution network planning considering bi-directional active-reactive power interaction through the sequential decomposition method

\textit{Case D:} Active distribution network planning considering bi-directional active-reactive power interaction using holistic optimization method

Cases A-C employ the sequential decomposition method to plan both the distribution network and V2GCS. In particular, Case C considers the bi-directional active-reactive power interaction between stations and multiple DGRs. However, Case A only considers the bi-directional interaction of active-reactive power of stations, and Case B solely accounts for the bi-directional active power interaction of stations and DGRs. Finally, Case D utilizes holistic optimization method to collaboratively plan ADN and V2GCS, taking into account the stations' bi-directional active-reactive interaction and DGRs. The associated costs and planning results are displayed in Table \ref{tab:comparison results} and Figure \ref{fig3}.

\begin{table}[htbp]
  \centering
  \caption{Annualized Construction and Operating Costs ($10^{4}$ CNY \textyen)}
  \label{tab:comparison results}
  \begin{tabular}{ccccc}
  \hline\hline
  Case No & \textit{Case A} & \textit{Case B} & \textit{Case C*} & \textit{Case D*}  \\ \hline
  Distribution network      &$\surd$ &$\surd$ &$\surd$ &$\surd$   \\ 
  V2GCS                     &$\surd$ &$\surd$ &$\surd$ &$\surd$   \\ 
  EV active power           &$\surd$ &$\surd$ &$\surd$ &$\surd$   \\
  EV reactive power         &$\surd$ &  ---   &$\surd$ &$\surd$   \\  
  DGRs                      &  ---   &$\surd$ &$\surd$ &$\surd$   \\ \hline
  Holistic optimization     &  ---   &  ---   &  ---   &$\surd$   \\ 
  Sequential decomposition  &$\surd$ &$\surd$ &$\surd$ &  ---     \\ \hline
  Network loss cost         & 460.38 & 440.64 &434.96 &472.58   \\ 
  Line construction cost    &59.58 &59.76  &59.76  &61.88     \\ 
  Investment cost of DGRs   &--- &11.10  &12.80  &17.36     \\ 
  O\&M cost of DGRs         &--- &5.52   &6.02  &11.34     \\ 
  Investment cost of V2GCS  &47.17 &47.17 &47.17  &47.17  \\ 
  O\&M cost of V2GCS        &14.10 &14.10 &14.10  &14.10  \\ \hline
  Total cost                & 581.23 & 578.29 &574.81  &624.43    \\
  \hline\hline
  \end{tabular}
  \begin{tablenotes}
  \footnotesize
  \raggedright
  \item{*} Compare optimization results for 6.80\% AEV penetration, with holistic method taking 10x longer than decomposition method (987.20s).
  \end{tablenotes}
\end{table}

\begin{figure}[htbp]
    \centering
    \includegraphics[width=0.45\textwidth]{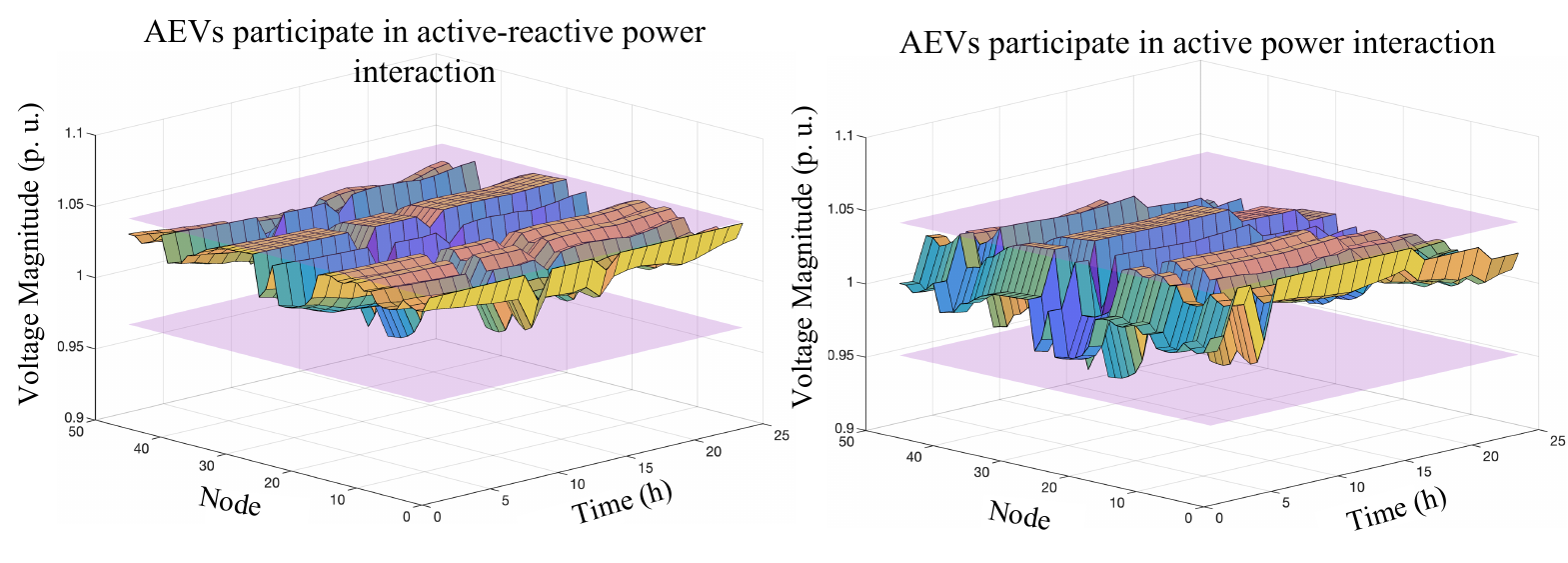}
    \caption{The effect of AEV's reactive power interaction on voltage distribution}\label{fig13}
\end{figure}

Case C has been found to be more cost-effective than Case A, which does not incorporate any DGRS. Specifically, the incorporation of DGRS reduces network loss cost by 5.52\%, as well as decreases line construction cost and total cost. This highlights how multiple DGRS can effectively lower the overall construction and O\&M costs of a power system. Furthermore, Case C was also found to be more economical than Case B, as it reduces network loss by 1.29\% and improves voltage quality. In Fig. \ref{fig13}, the effect of AEV's reactive power interaction on voltage distribution of the power system is depicted. The results indicate that the extreme difference in voltage was reduced by 17.61\%, while the variance dropped by 28.64\%. It is important to note that, when compared to bi-directional active power interaction, the use of bi-directional active-reactive power interaction results in even greater reduction in voltage fluctuations, hence further reducing the construction and O\&M cost.

Cases C and D share identical modeling content but adopt different solution methods. The employment of decomposition method in Case C yielded a result with a 0.00\% gap within 98.72 seconds, while the holistic method took 10x longer to resolve and generate as observed in Case D. Such observations highlighted that the expenses incurred by the planned solutions outweighed those of the holistic method, leading to an overall escalation of 8.63\% in construction and O\&M cost. These revelations exemplified the supremacy of the decomposition approach, which had the capacity to deliver better optimal planning results within a reduced timeframe.

\begin{figure}[htbp]
    \centering
    \includegraphics[width=0.45\textwidth]{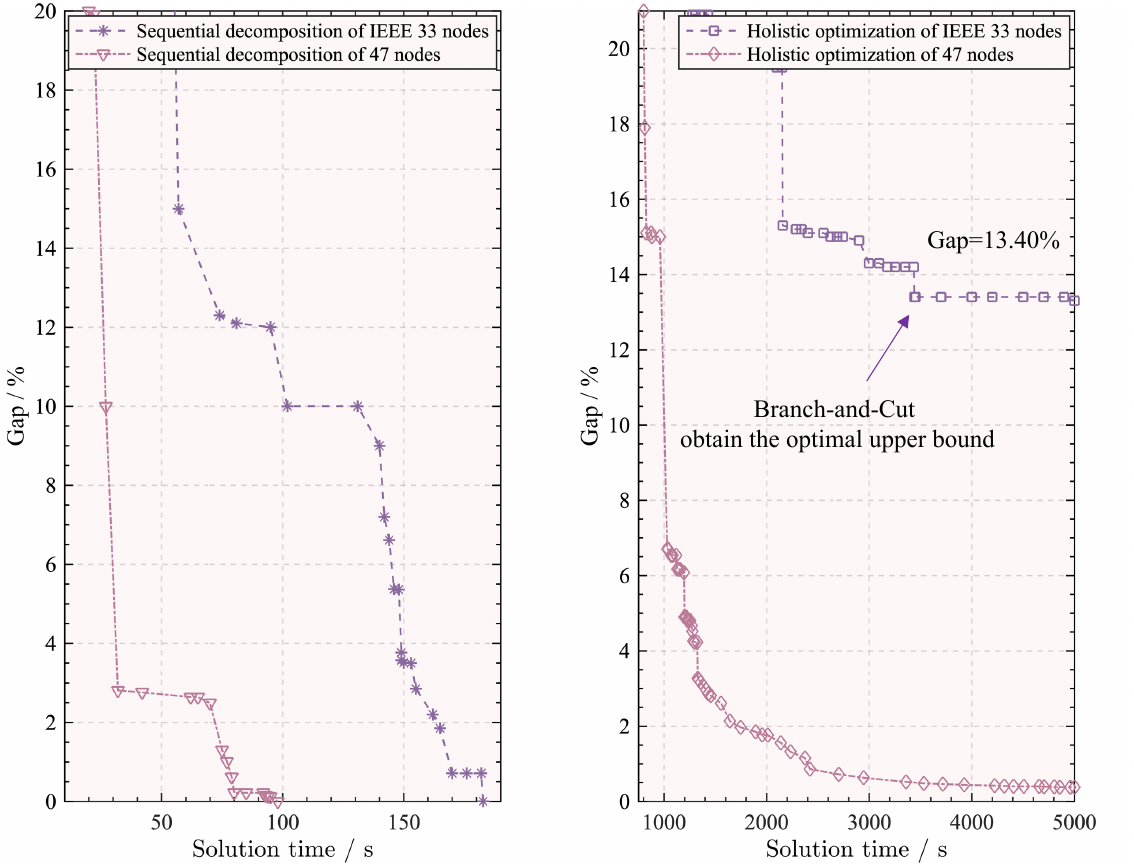}
    \caption{Solution gap of sequential decomposition and holistic method}\label{fig4}
\end{figure}

Continuing with the holistic method from Case D, a solution gap of 0.39\% was obtained after 5000 seconds of computation. Remarkably, the distribution grid construction plan, V2GCS retrofitting and construction options, and various DGR siting and sizing recommended by Case D were found to be completely identical to those obtained by the planning results in Case C that only took 98.72 seconds. Validation of both methods on the IEEE 33 nodes system revealed that the sequential decomposition method achieved a solution gap of 0.00\% in 182.72 seconds, while the holistic approach failed to produce an optimal solution even after 5000 seconds. The solution gaps for both methods are depicted in Fig. \ref{fig4}. Moreover, the decomposition method surpassed the holistic approach, resulting in a 3.54\% reduction in overall construction and O\&M costs in IEEE 33 nodes system.

\subsection{Load Profile Analysis of V2GCS}

\begin{figure}[htbp]
    \raggedright
    \includegraphics[width=0.47\textwidth,height=182pt]{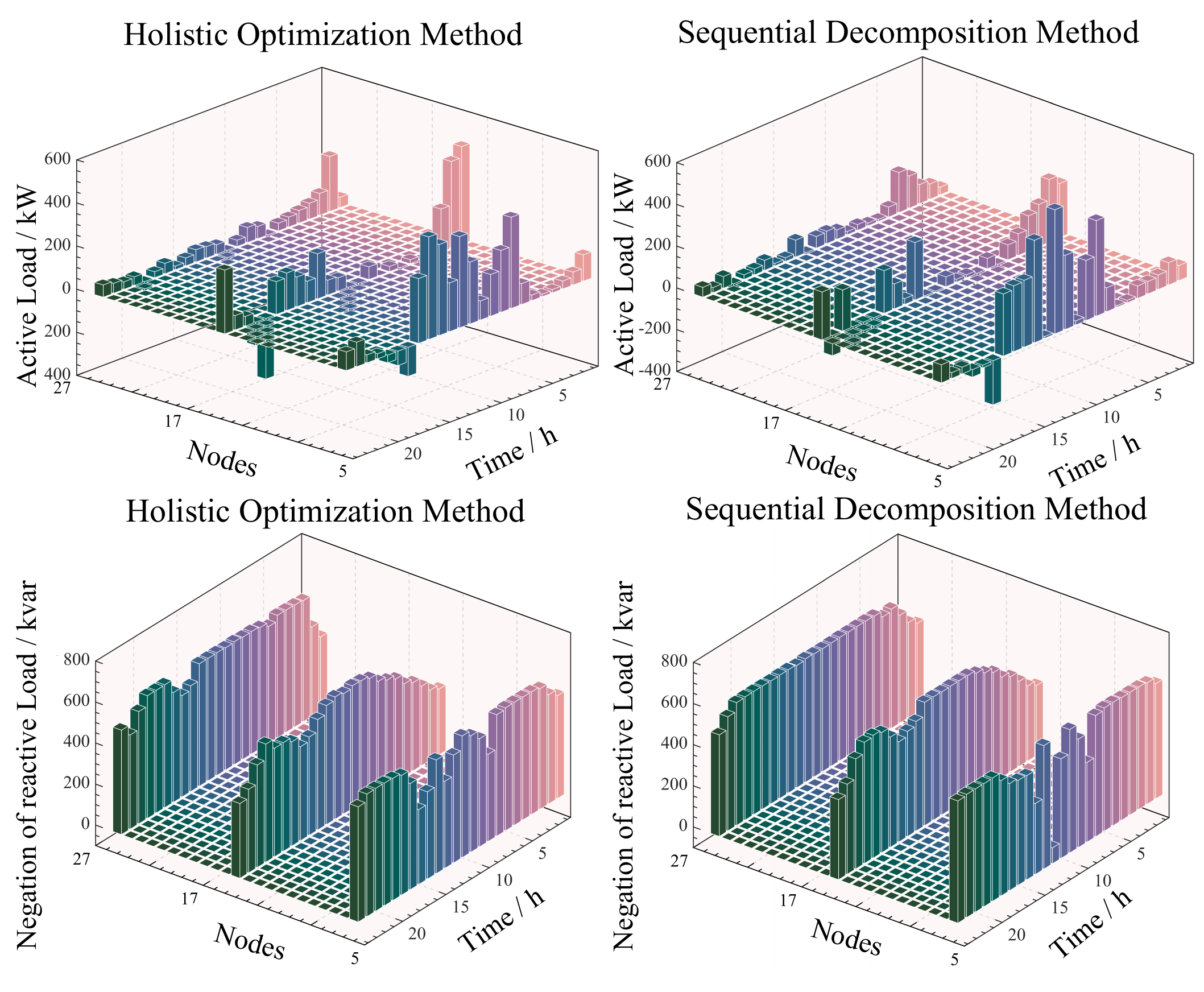}
    \caption{Active and reactive load of V2GCS}\label{fig5}
\end{figure}

At an AEV penetration rate of 6.80\%, both the decomposition and holistic methods in Case C and Case D yielded identical planning outcomes despite the stark contrast in computational times. The former method took a mere 98.72 seconds, while the latter consumed 5000 seconds. The common strategy involved constructing a new V2GCS at Node 5 and retrofitting the existing EV charging stations at Node 17 and Node 27 into V2GCS. As illustrated in Fig. \ref{fig5}, the V2GCS load was distributed among residential, office, and industrial areas. Notably, no retrofit or new construction of V2GCS was deemed necessary for the commercial area due to the paucity of AEVs that required slow-charging services.

Despite the regulation efforts implemented by the information processing center, the charging load of the stations remains primarily concentrated during the peak daytime period owing to the commuting behavior of AEVs in office area. During the evening, the charging process will transition to discharging power back to the grid. However, in residential and industrial areas, a feasible solution to shift peak loads from high-demand periods to low-demand periods during the night is by guiding AEVs' charging and discharging behavior, with the extra benefit of discharging power back to the grid during peak hours.

The results presented in Fig. \ref{fig5} demonstrate that the two methods exhibit almost identical active and reactive station features across various areas, highlighting the approximately equivalent planning outcomes observed in Case C and Case D. Therefore, the sequential decomposition method provides a high-quality solution for addressing large-scale problems that may be challenging in attaining optimal outcomes using the holistic optimization method.

\subsection{Analysis of Results at Different AEV Penetration}

Based on the analysis presented in Fig. \ref{fig4}, our study reveals that the solution complexity of the IEEE 33 nodes distribution network exceeds that of the Longgang district 47 nodes distribution grid, and that the optimization gap exhibits a slower convergence rate. To investigate the impact of varying AEV penetration rates on both the distribution network topology and V2GCS, we employ the IEEE 33 nodes network as a case study in this section. Moreover, we consider that a global optimal solution has not been attained if the optimization process has lasted for 5000 seconds and the optimization gap has not decreased to less than 5\%.

\begin{figure}[htbp]
    \centering
    \includegraphics[width=0.48\textwidth]{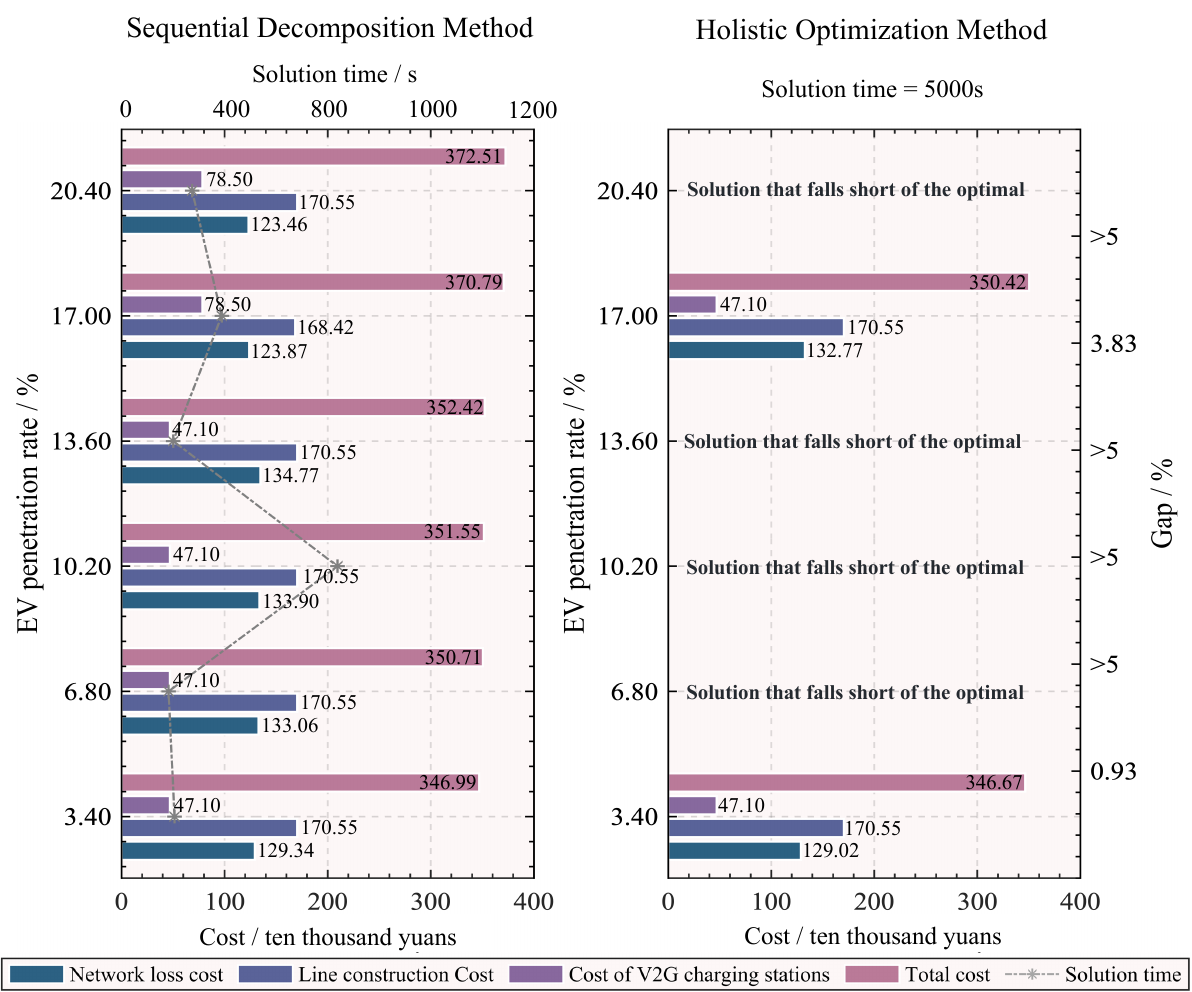}
    \caption{Planning costs for different AEV penetration rates}\label{fig6}
\end{figure}

Fig. \ref{fig6} depicts the construction and O\&M cost, as well as solution times, of two different methods under varying AEV penetrations. Our analysis indicates that the sequential decomposition method yields planning solutions in a relatively short time frame, regardless of the AEV penetration. Conversely, the holistic optimization method takes a longer time to converge and does not attain the global optimum at AEV penetrations of 6.80\%, 10.20\%, 13.60\%, and 20.40\%. Further scrutiny of the planning costs for the sequential decomposition method reveals a positive correlation with an increase in AEV penetration, while the solution time displays no discernable pattern.

At an AEV penetration of 3.4\%, the sequential decomposition method produced the same planning solution as the holistic optimization method, but in a significantly shorter time of 208 seconds compared to 5000 seconds for the latter. However, at an AEV penetration of 17.00\%, the planning cost for the sequential decomposition method exceeded that of the holistic optimization method. In scenarios with high AEV penetration, the coupling effect between $x$ and $y$ is strengthened, and optimizing AEV behavior may significantly impact global optimal solutions. Prioritizing AEV behavior optimization may require additional V2GCS construction to minimize network loss and line construction costs, leading to higher overall planning costs. In such cases, local optimal solutions may diverge from global optimal solutions, rendering the sequential decomposition method suitable for planning problems with low AEV penetration.
\subsection{Numerical Validation of the Solution Existence}

Taking the planning results obtained with the proposed sequential decomposition method as an example, the distribution grid is constructed according to the layout depicted in Fig. \ref{fig3} for Case C, with the corresponding DGRs and V2GCS invested at certain locations. Even in the worst case scenario, where all the AEVs in the V2GCS are simultaneously charging at their maximum power, the load profile $\overline{y^{*}}$ still complies with the constraints \eqref{eq59} of the ADN, ensuring the secure and stable operation of the power system, as well as the solution's existence of the decomposed model.

\begin{figure}[htbp]
    \flushleft
    \includegraphics[width=0.45\textwidth]{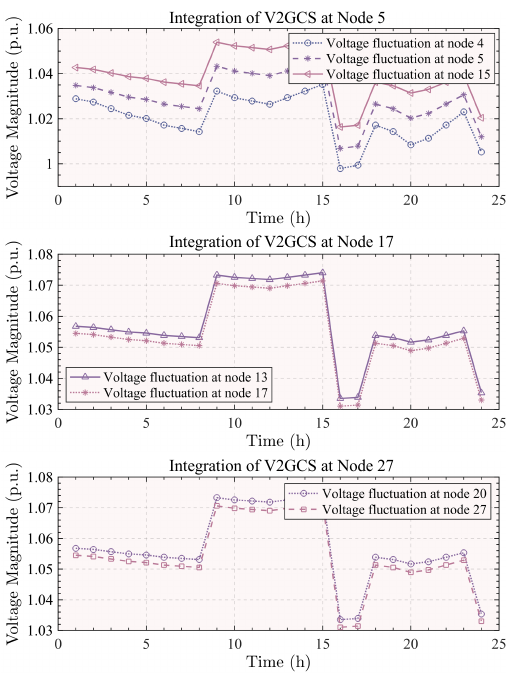}
    \caption{Voltage fluctuation of V2GCS and neighboring nodes}\label{fig10}
\end{figure}

In Case C, V2GCS installations are deployed at specific nodes within the residential area (Node 27), office area (Node 12), and industrial area (Node 17). The primary analysis focuses on the power flow of the interconnected lines associated with the V2GCS installation nodes, as well as the voltage distribution at these nodes and their neighboring nodes. The distribution power flows are within the line capacity, while the voltage distribution, as depicted in Fig. \ref{fig10}, also meets the system's requirement. Consequently, under any scenario of AEVs charging or discharging, the feasibility of the sequential decomposition method is ensured.
\section{Conclusion}

In this study, we introduced a sequential decomposition method based on MILP and MISOCP to collaboratively plan ADN and V2GCS with consideration of AEV characteristics in different regions. Our planning approach prioritizes fulfilling the energy demands of AEV customers, while also accounting for their behavioral characteristics. This method is achieved through the integration of almost all regulating devices within the power grid, as well as a future distribution network background of the widespread adoption of autonomous driving technology.

Our proposed decomposition method effectively and efficiently solves large-scale planning problems. Holistic optimization solutions may encounter extreme slowness or may even be unsolvable when the problem scale is large. In contrast, our sequential decomposition method can efficiently obtain planning results while satisfying accuracy requirements at low AEV penetration.

Furthermore, our results demonstrate that implementing multiple DGRs can reduce planning costs. When planned rationally as a new energy source, retrofitted V2GCS with bi-directional active-reactive power interaction has the potential to not only decrease construction and O\&M costs but also improve power grid stability.

\section*{Appendix}

\begin{figure*}[htbp]
    \centering
    \includegraphics[width=0.95\textwidth]{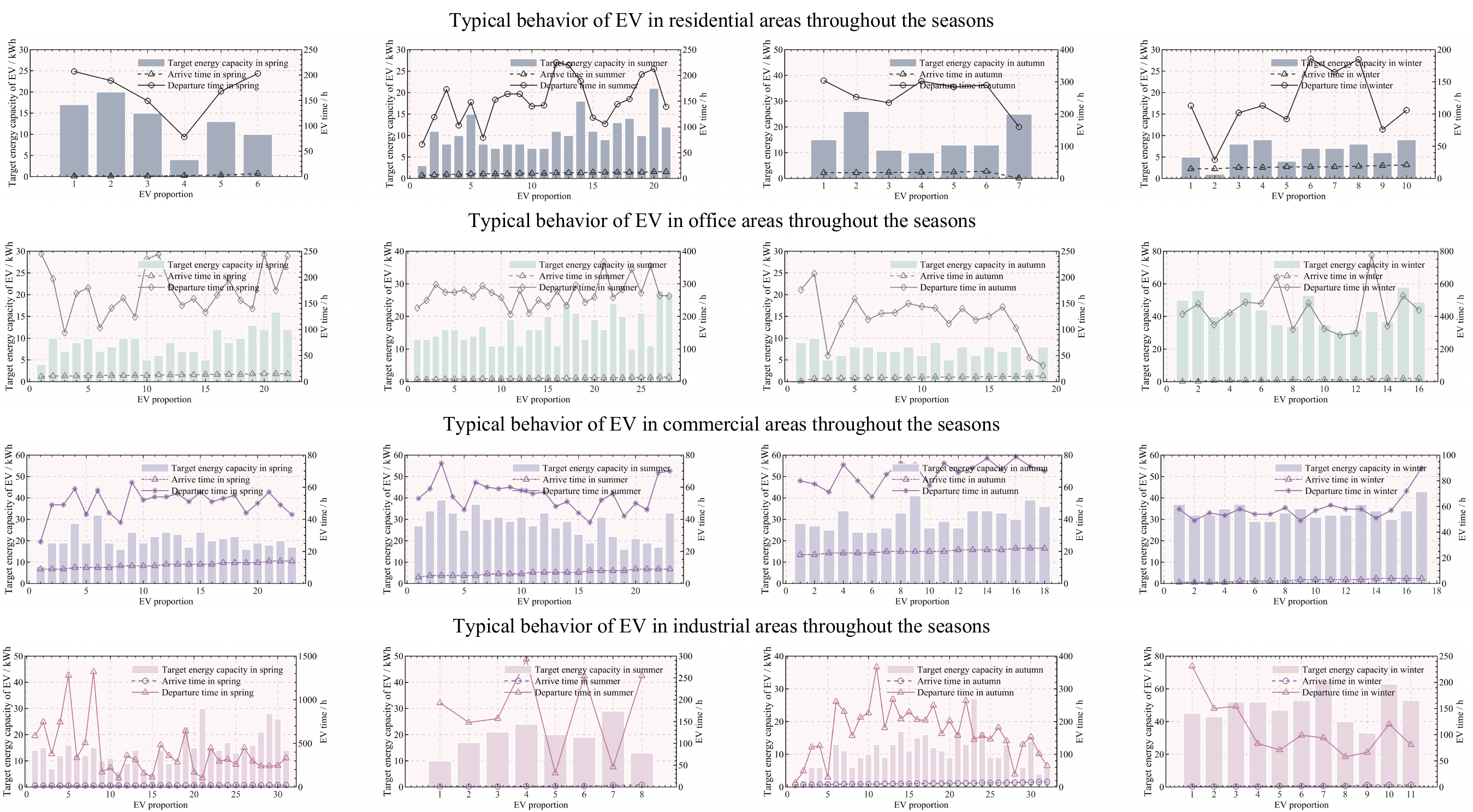}
    \caption{Typical behavior of EV in Longgang District, Shenzhen, China}\label{fig7}
\end{figure*}

Fig. \ref{fig7} illustrates typical EV behavior in Shenzhen, China, encompassing arrival times, departure times, and target energy capacities.

\begin{figure}[htbp]
    \centering
    \includegraphics[width=0.45\textwidth]{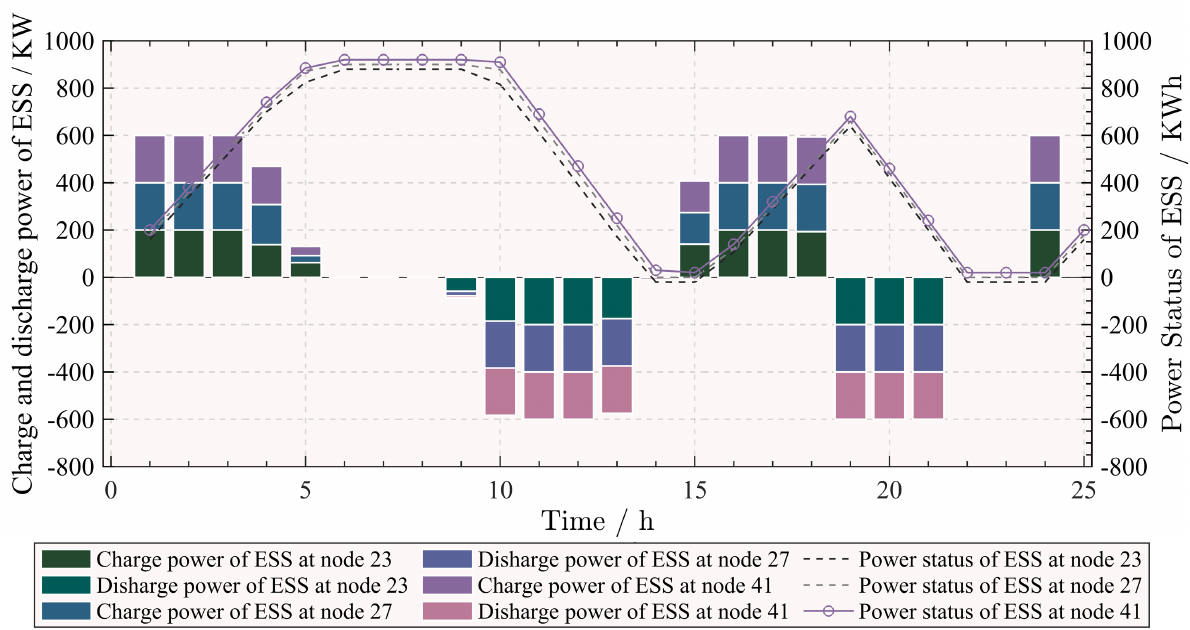}
    \caption{Operation of ESS in case C}\label{fig8}
\end{figure}

Fig. \ref{fig8} displays the ESS operation in case C, illustrating the power flow between the grid and ESS.

\begin{figure}[htbp]
    \centering
    \includegraphics[width=0.45\textwidth]{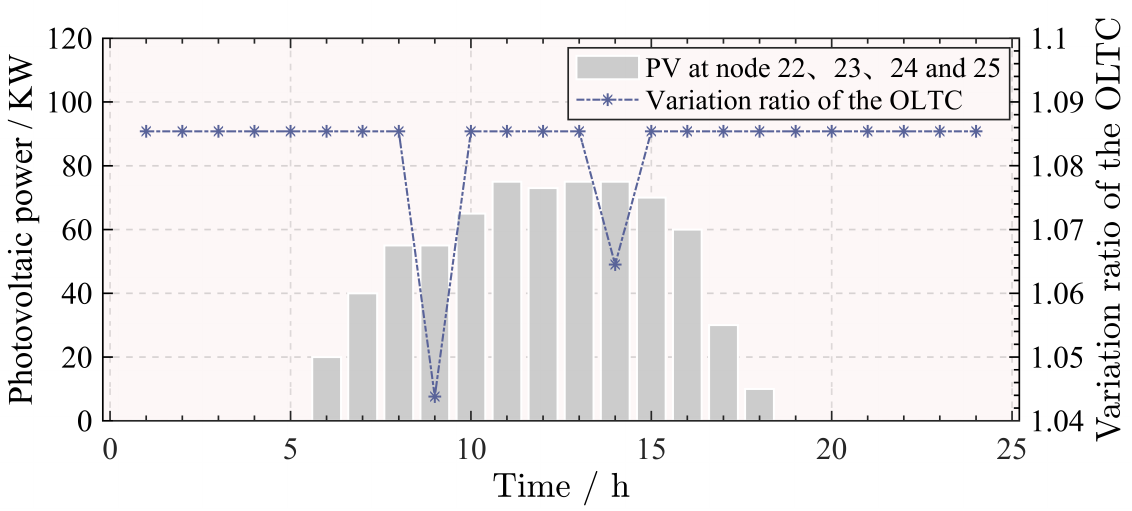}
    \caption{Operation of PV and OLTC in case C}\label{fig9}
\end{figure}

Fig. \ref{fig9} showcases the operation of PV and OLTC in case C, presenting the power output of the PV and the corresponding tap position of OLTC.

\section*{Acknowledgment}

The authors would like to express their gratitude to Naghmash Ali for his valuable comments and feedback, which greatly contributed to the improvement of this paper.

\bibliographystyle{IEEEtran}
\small\bibliography{reference}

\begin{IEEEbiography}[{\includegraphics[width=1in,height=1.25in,clip,keepaspectratio]{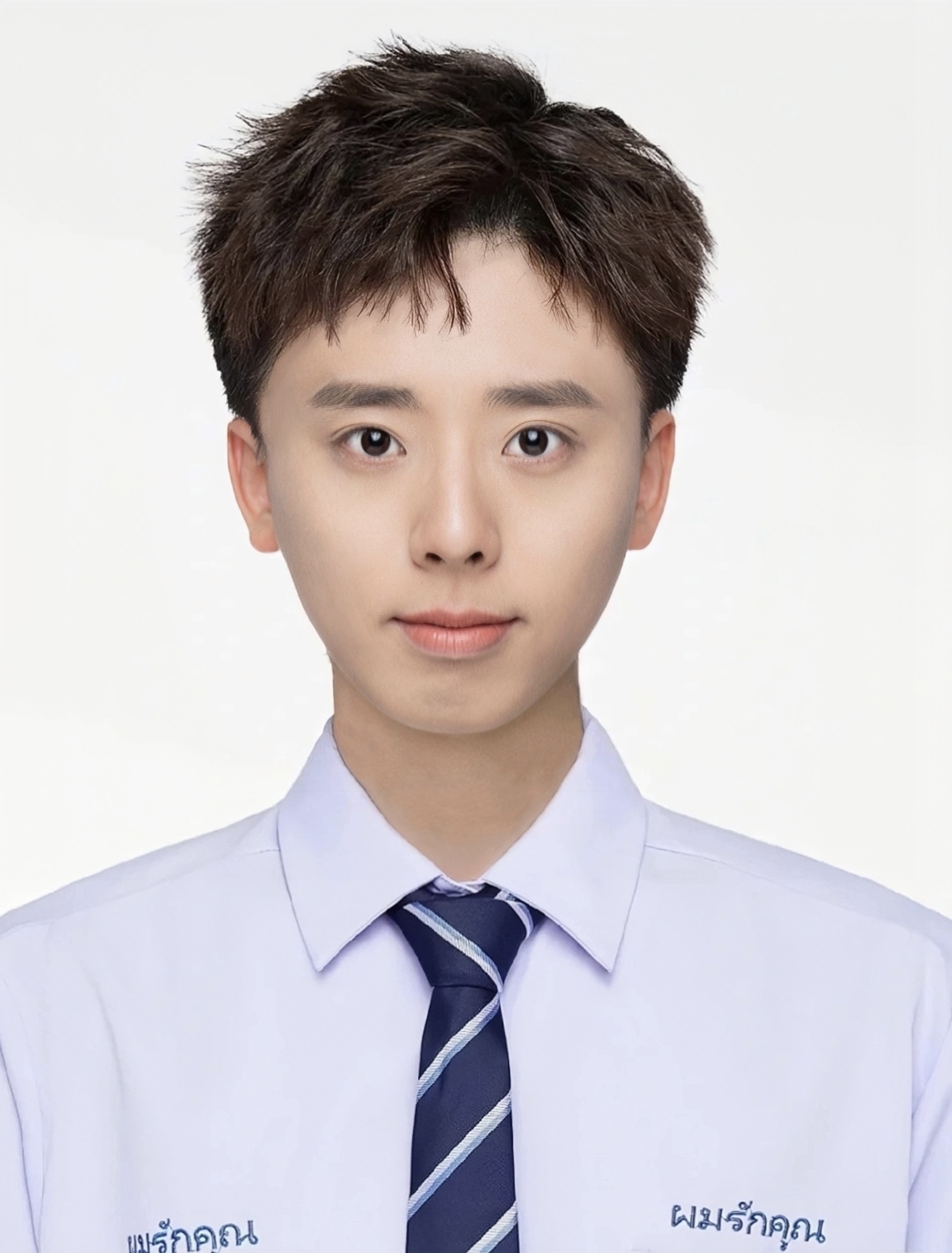}}]{Yongheng Wang}
(S’22) received his Bachelor's degree in 2021 from the Department of Electrical Engineering at South China University of Technology. He is currently pursuing a Master's degree at Tsinghua Shenzhen Institute Graduate School, Tsinghua University. His research interests encompass the intersection of transportation and power systems, with a particular focus on leveraging mathematical optimization in the smart grid.
\end{IEEEbiography}

\begin{IEEEbiography}[{\includegraphics[width=1in,height=1.25in,clip,keepaspectratio]{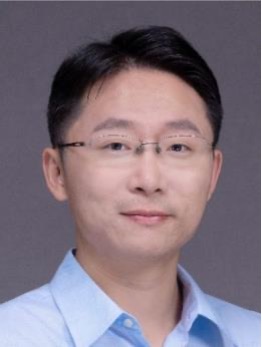}}]{Xinwei Shen}
(M'12–SM'21) graduated from the Dept. of Electrical Engineering, Tsinghua University, Beijing, China in 2010 and 2016, respectively, with B.Eng. and Ph.D. degrees. He was a visiting scholar at IIT, U.C.Berkeley and Univ. of Macau in 2014, 2017 and 2021. He is now an Assistant Professor in Tsinghua Shenzhen International Graduate School, Tsinghua University. In 2020, he was awarded the Young Elite Scientists Sponsorship Program by Chinese Society for Electrical Engineering (CSEE). His research interests include energy internet, integrated energy system/power distribution system optimization with energy storage and renewables. He is currently an IEEE senior member, co-chair of IEEE PES Working Group on Integrated Energy System/Multi-Energy Network Modeling and Planning and officer of PES Energy Internet Coordinating Committee (EICC). He is a subject editor of CSEE Journal of Power and Energy Systems and young editorial board member of Applied Energy.
\end{IEEEbiography}

\begin{IEEEbiography}[{\includegraphics[width=1in,height=1.25in,clip,keepaspectratio]{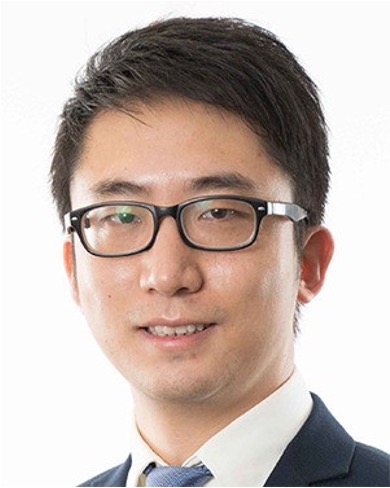}}]{Yan Xu}
(S’10-M’13-SM’19) received the B.E. and M.E degrees from South China University of Technology, Guangzhou,
China in 2008 and 2011, respectively, and the Ph.D. degree from The University of Newcastle, Australia, in 2013. He conducted postdoctoral research with the University of Sydney Postdoctoral Fellowship, and then joined Nanyang Technological University (NTU) with The Nanyang Assistant Professorship. He is now an Associate Professor at School of Electrical and Electronic Engineering and a Cluster Director at Energy Research Institute @NTU(ERI@N). His research interests include power system stability and control, microgrid, and data-analytics for smart grid applications. Dr Xu's professional service roles include Associate Editor for several international journals including IEEE Trans. Smart Grid and IEEE Trans. Power Systems, Chairman of the IEEE Power \& Energy Society (PES) Singapore Chapter (2021 to 2022) and General Co-Chair of the 11th IEEE ISGT-Asia Conference, Nov. 2022.
\end{IEEEbiography}

\end{document}